# Concurrent magneto-optical imaging and magneto-transport readout of electrical switching of insulating antiferromagnetic thin films

F. Schreiber,[1] L. Baldrati,[1, a)] C. Schmitt,[1] R. Ramos,[2] E. Saitoh,[2,3,4,5,6] R. Lebrun,[1,7, b)] and M. Kläui[1,8]

[1]Institute of Physics, Johannes Gutenberg-University Mainz, 55128 Mainz, Germany

[2]WPI-Advanced Institute for Materials Research, Tohoku University, Sendai 980-8577, Japan

[3]Institute for Materials Research, Tohoku University, Sendai 980-8577, Japan

[4]Advanced Science Research Center, Japan Atomic Energy Agency, Tokai 319-1195, Japan

[5]Center for Spintronics Research Network, Tohoku University, Sendai 980-8577, Japan

[6]Department of Applied Physics, The University of Tokyo, Tokyo 113-8656, Japan

[7]Unité Mixte de Physique CNRS, Thales, Univ. Paris-Sud, Université Paris-Saclay, Palaiseau 91767, France

[8]Graduate School of Excellence Materials Science in Mainz, 55128 Mainz, Germany

[a)] lbaldrat@uni-mainz.de

[b)] romain.lebrun@cnrs-thales.fr

## ABSTRACT

We demonstrate stable and reversible current induced switching of large-area (> 100 µm²) antiferromagnetic domains in NiO/Pt by performing concurrent transport and magneto-optical imaging measurements in an adapted Kerr microscope. By correlating the magnetic images of the antiferromagnetic domain changes and magneto-transport signal response in these current-induced switching experiments, we disentangle magnetic and non-magnetic contributions to the transport signal. Our table-top approach establishes a robust procedure to subtract the non-magnetic contributions in the transport signal and extract the spin-Hall magnetoresistance response associated with the switching of the antiferromagnetic domains enabling one to deduce details of the antiferromagnetic switching from simple transport measurements.

**MANUSCRIPT**

Antiferromagnetic materials (AFMs) potentially offer outstanding properties such as high bit packing density thanks to the absence of stray fields, insensitivity to magnetic fields, operation with THz speeds and thus, a promising platform for future spintronic devices.[1] One key requirement in view of applications is the ability to electrically write and read defined states of the orientation of the AFM Néel order $\boldsymbol{n}$. Recently, evidence for electrical switching was found and corresponding switching mechanisms were proposed both in metallic AFMs[2,3] and insulating AFM/heavy metal bilayers.[4–10] For AFM insulators, the electrical readout mechanism relies on the transverse spin Hall magnetoresistance (SMR) generated e.g. in an adjacent Pt layer where the signal depends on $n_x * n_y$, the product of the Néel vector components.[8,11–14] However, recent reports suggest that a commonly observed triangular-shaped feature in the electrical signal is the result of non-magnetic contributions such as electromigration, thermal effects or other artifacts in the heavy metal layer.[7–9,15–17] In addition to the triangular-shaped non-magnetic contributions, a steplike magnetic SMR contribution was reported in NiO[7] and α-Fe$_2$O$_3$.[9] These results have generated a debate on the magnitude of the magnetic contributions, the reliability of the SMR readout mechanism and even on the general possibility of switching AFM domains electrically.[14,15,17] To answer these questions, the interplay of magnetic and non-magnetic contributions has to be clarified by simultaneously quantifying the electrical transport signal response and the changes of the AFM spin structure determined independently. To date, AFM domain imaging has been predominantly performed by utilizing the x-ray magnetic linear dichroism effect (XMLD) in photoemission electron microscopy (PEEM) setups.[5,7,18,19] Thus, the direct comparison of the magnetic imaging of antiferromagnetic domains and the electrical signal has been hampered by the difficulty to combine low-noise electrical detection (necessary for SMR) and synchrotron-based imaging.[5,7] Recent reports on lab-based imaging techniques that potentially simplify such an attempt include spin Seebeck microscopy, which still requires a sophisticated setup including a sufficiently high-power laser.[6] Alternatively, an optical imaging technique utilizing the magneto-optical birefringence effect was reported for imaging the AFM domain structure of NiO and CoO with a commercial Kerr microscope.[20,21]

In order to clarify the existence of current-induced antiferromagnetic switching and correlate the electrical transport signals and the imaging results, one needs to carry out both concurrently. In this work, we take advantage of simple optical imaging of the antiferromagnetic domains to firstly unambiguously demonstrate the actual large-area electrical switching of the Néel order in epitaxial NiO thin films and secondly directly correlate the imaged antiferromagnetic domains to the electrical response. This allows us to clarify the possibility to read and write the Néel order orientation in AFMs electrically. Moreover, we distinguish magnetic and non-magnetic contributions to the transverse resistance and provide a simple background subtraction procedure to robustly identify the magnetic contribution to the electrical transport signal.

NiO is a well-known collinear antiferromagnet with a Néel temperature of 523 K in the bulk[22] that exhibits, in combination with a Pt layer, negative SMR.[12,23] It constitutes an ideal platform to simultaneously achieve electrical switching due to its low anisotropy[12,13] and magneto-optical imaging due to a large magneto-optic birefringent coefficient[20] at room temperature. To study current induced switching in the insulating NiO, we fabricated NiO(001)(10nm)/Pt(2nm) bilayers by epitaxial growth on MgO(001). Apart from a slight increase of the oxygen flow to improve the growth conditions, we used the growth procedure described in Ref. [7] where we verified the antiferromagnetic ordering of the spins by observation of x-ray magnetic linear dichroism (XMLD) and absence of x-ray magnetic circular dichroism (XMCD). In order to apply current pulses and measure the transverse resistance, we patterned eight-terminal devices with optical lithography and subsequent Ar ion etching. In these samples it was possible to image the antiferromagnetic domains according to the technique shown in Ref. [20]. Fig. 1(a) depicts the device layout and measurement scheme. We applied 0.1 ms-long current pulses, called $I_{P\ +45°}$ and $I_{P\ -45°}$, along the [110] and [1$\bar{1}$0] crystallographic directions of the NiO, respectively. Subsequently, the transverse voltage $V_{Read}$ was measured along the [010] direction with a reading current density $j_{Read} \sim 10^9\ A\ m^{-2}$ along the $\langle 100 \rangle$ directions. Between the application of a write pulse and the reading operation we set a delay of 10 s to mitigate the influence of Joule heating on the measured signal, taking advantage of an automated matrix switch system to route the contacts. We calculate the transverse resistance as $R_{transv} = [V(I^+_{Read})/I^+_{Read} + V(I^-_{Read})/I^-_{Read}]/2$, with positive and negative current readings to reduce the thermal contribution to the transverse voltage $V$.

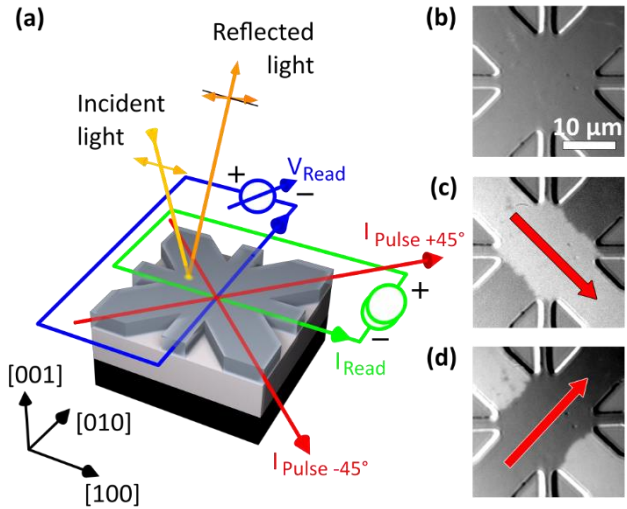

Fig. 1: (a) Device layout and measurement scheme relative to the crystallographic directions of the 10 nm NiO film (intermediate layer). The polarity of electrical readout measurements is indicated by the current source and voltmeter. In such samples, the antiferromagnetic domains can be imaged by utilizing the magneto-optic birefringence effect. Examples of domain images show (b) the virgin state of a sample and the domain structure after $j_P = 1.5 \times 10^{12}\ A\ m^{-2}$ writing pulses in $-45°$ direction (c) and $+45°$ direction (d) as indicated by the red arrows. The scalebar provides the scale of all domain images.

We performed optical imaging of the AFM domain state with a commercial magneto-optical Kerr microscope, following the procedure described by Ref. [20]. The magneto-optical birefringence effect results in opposite polarization rotation of the polarized light reflected from regions with orthogonal Néel vectors. Thus, the difference image obtained for two analyzer directions with reverse angles from the extinction position yields a net contrast corresponding to the in-plane projection of the Néel vector. The observed two-level contrast is attributed to the two orthogonal projections of T-domains which are known to show fourfold in-plane symmetry in NiO thin films, with in plane projections of the Néel order along the ⟨110⟩ directions.[20,24] Note that we apply the measurement current along ⟨100⟩ such that the in-plane projection of the anisotropy axis is $\pm45°$ with respect to the current direction, which is expected to maximize the difference in the transverse SMR amplitude ($n_x * n_y$) between the two stable domain states. The virgin domain state of a four-channel 10 µm Hall star device is visible in Fig. 1(b) and is largely single domain. Figs. 1(c),(d) show the domain state after the application of a current pulse with a current density of $j_P = 1.5 \times 10^{12} \, A \, m^{-2}$. The pulse along the $-45°$ direction reverses the contrast along the whole 10 µm channel (Fig. 1(c)), while a subsequent pulse in $+45°$ direction switches the central area of the device back to the initial state (Fig. 1(d)). The application of pulses with alternating current paths results in a full and reversible switching between the two magnetic states shown here. Note that the switching occurs for a large area of hundreds of µm² and saturates indicating full switching. Upon decreasing the pulse current density, we see a decrease in the fraction of switched area. The observed changes of the domain structure are strongly coupled to the pulse direction and thereby unambiguously confirm the possibility of writing the Néel order orientation and AFM domain state electrically. We found from such optical images that the current induced magnetic configuration is stable over at least 3 weeks without any relaxation, such that the written domain states serve two basic requirements for the application in spintronic devices: reproducible switching between two defined states and long term stability of the written state. However, a reliable electrical readout is necessary in view of applications, and it is thus a key requirement to eliminate the non-magnetic contributions to the electrical signal. We can single out the magnetic contribution by a comparison to the changes observed by imaging. The impact of a current pulse on the domain structure can be quantified by calculating the difference of initial and final domain images. The analysis of the difference image yields the number of pixels corresponding to the switched area after the application of the current pulse (see supplementary material for further detail).

In Fig. 2(a) we show a comparison between the switched area measured in pixels and the corresponding as-measured electrical signal obtained during cycles of 5 pulses in alternating directions at a current density significantly above switching threshold. Firstly one sees that the switched area and the as-measured electrical signal at this current density deviate from each other: by applying many identical current pulses, we observe saturation to a constant domain state after about the fourth pulse (orange up triangles), whereas the electrical signal increases further with a linear slope (blue discs). The behavior reverses upon rotation of the pulse direction by 90 degrees, thus, showing in the switched area calculated from the optical images, a reversible and quasi-steplike switching of the domain state and in the as-measured electrical signal a triangular-shaped behavior with a larger increase for the first pulses than for later pulses. This comparison confirms the non-magnetic origin of the triangular-shaped component of the as-measured electrical signal, in line with Refs. [7–9,15–17]. From this point on, we subtract the triangular-shaped contribution from the electrical signal based on the slope between pulse 4 and 5 of each direction, where we observed saturation of the domain state. The resulting data (purple squares) shows a striking similarity to the switched area measured optically (orange up triangles) and thereby confirms that the larger initial increase in the electrical signal corresponds to the presence of a significant magnetic-related (SMR) contribution as reported in Ref. [7,9].

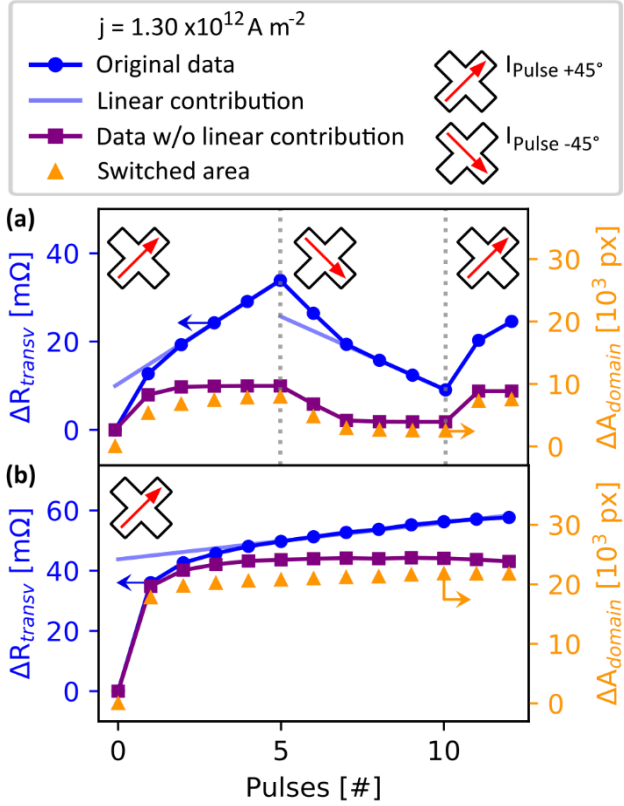

Fig. 2: Comparison of current-induced changes of the domain structure measured by direct imaging (orange up triangles) with the corresponding electrical signal (blue discs) and the signal after subtraction of a linear contribution (purple squares) as a function of the pulse orientation and number. The orange and purple data are plotted with an offset for clarity. The pulse current density was $j_P = 1.30 \times 10^{12} \, A \, m^{-2}$ with the pulse direction as indicated by the insets. (a) Alternating switching after a $-45°$ initialization pulse of $j_P = 1.35 \times 10^{12} \, A \, m^{-2}$ and (b) a single switching measurement after a $-45°$ initialization pulse of $j_P = 1.50 \times 10^{12} \, A \, m^{-2}$. The difference of the initial states results in a difference of the magnitude of $\Delta R_{transv}$ in these two measurements.

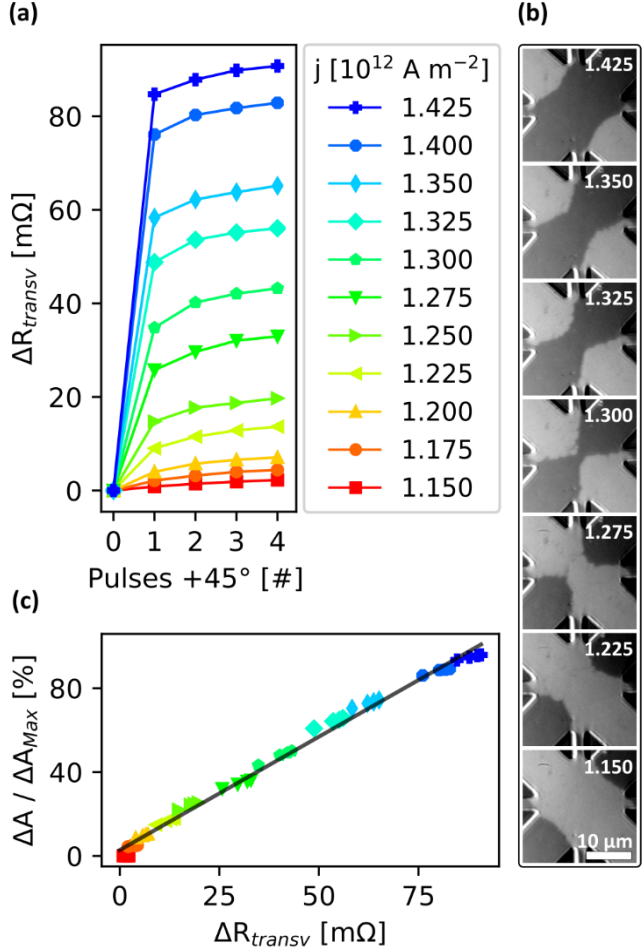

Fig. 3: (a) Electrical signal after subtraction of a linear contribution for the accessible current densities. The system was always reset to the same initial state ($-45°$ pulse, $j_P = 1.5 \times 10^{12}\ A\ m^{-2}$). (b) Domain structure for different current densities after application of 15 pulses in the $+45°$ direction. The initial domain state compares to Fig. 1(c). The scalebar provides the scale of all domain images. (c) Correlation of electrical signal and switching fraction calculated from the imaging.

After assessing how to subtract the non-magnetic contributions, we directly compare the electrical signal to the imaged switched area as a function of the pulse current density. We induced a saturated monodomain state of the central area by applying an initial $-45°$ pulse of $j_P = 1.5 \times 10^{12}\ A\ m^{-2}$ (compare Fig. 1(c)) and subsequently applied 15 current pulses in the $+45°$ direction. Below the threshold pulse current density of $j_P = 1.15 \times 10^{12}\ A\ m^{-2}$ we do not observe changes of the domain structure. For an intermediate current density of $j_P = 1.30 \times 10^{12} \, A \, m^{-2}$ the comparison of switched area (orange up triangles, see also supplementary material, Fig. S1) and electrical measurement (blue discs) is presented in Fig. 2(b). The behavior observed here is similar for all current densities above the threshold: the changes of the domain structure (switched area) are large after the first pulse, quickly saturate and result in a final domain state. The size of the total switched area monotonically increases with the pulse current density (compare Fig. 3(b)). The transverse resistance signal features similar changes, which are typically large at the beginning and saturate towards a quasi-linear, current dependent slope after several pulses (as-measured data in supplementary materials, Fig. S2). Next, we check the applicability of our approach to identify the magnetic contribution to the transport signal. For $j_P < 1.45 \times 10^{12} \, A \, m^{-2}$ we can accurately subtract the non-magnetic contribution by a linear fit to the data from pulse 4 onwards. The thus treated electrical data of the first 4 pulses for all the accessible current densities is included in Fig. 3(a), while Fig. 3(b) shows the images of the final domain states taken at different pulse current densities. To check how good this procedure is, we compare the amplitude of the switching observed in the post-treated electrical signal and the fraction of domain switching obtained by imaging the switched area. We find that for the current densities studied, both increase linearly with current in the same manner. This correlation is quantified in Fig. 3(c) where we plot the relative switched area as a function of the electrical signal amplitude. The linear fit is excellent ($R^2 = 0.997$) indicating that the treated electrical signal is a measure for the magnetic switched area and we conclude that our procedure efficiently subtracts the non-magnetic contributions to the electrical signals. This simple linear subtraction holds within the timescale of our experiments and over a significant range of current densities.

To summarize, we achieve current-induced electrical writing and reading of insulating antiferromagnetic materials. By comparing the electrical signal and the optical imaging of the magnetic switching, we demonstrate that the transverse resistance is composed of a magnetic quasi-steplike contribution, which is linearly proportional to the switched area, and a non-magnetic triangular-shaped contribution.[7,9] Based on this comparison, we propose a procedure to robustly identify the magnetic and non-magnetic contributions to the electrical transport signal from the behavior of the transverse resistance as a function of the number of current pulses. Together, our results unambiguously show the possibility of electrical reading and writing of antiferromagnetic insulators by subtracting non-magnetic contributions to the electrical signal. This motivates additional research to enable using AFMs in applications where electrical reading and writing are key and can in parallel motivate the development of antiferromagnetic opto-spintronic devices,[25] thanks to the large magneto-optical response observed in NiO.

See the supplementary material for further details on the procedure used to extract the domain areas and the as-measured electrical signal of the single switching measurements.

## ACKNOWLEDGMENTS

The authors thank H. Meer for skillful technical assistance and insightful discussions. L.B acknowledges the European Union's Horizon 2020 research and innovation program under the Marie Skłodowska-Curie grant agreements ARTES number 793159. L.B., R.L., and M.K. acknowledge support from the Graduate School of Excellence Materials Science in Mainz (MAINZ) DFG 266, the DAAD (Spintronics network, Project No. 57334897) and all groups from Mainz acknowledge that this work was funded by the Deutsche Forschungsgemeinschaft (DFG, German Research Foundation) - TRR 173 – 268565370 (projects A01, A03, A11, B02, and B12). R.L. and M.K. acknowledge financial support from the Horizon 2020 Framework Programme of the European Commission under FET-Open grant agreement no. 863155 (s-Nebula). This work was also supported by ERATO "Spin Quantum Rectification Project" (Grant No. JPMJER1402) and the Grant-in-Aid for Scientific Research on Innovative Area,

"Nano Spin Conversion Science" (Grant No. JP26103005), Grant-in-Aid for Scientific Research (C) (Grant No. JP20K05297) from JSPS KAKENHI, Japan. R.L. acknowledges the European Union's Horizon 2020 research and innovation program under the Marie Skłodowska-Curie grant agreement FAST number 752195.

The data that support the findings of this study will be openly available in Zenodo upon acceptance of the manuscript.

## REFERENCES

[1] V. Baltz, A. Manchon, M. Tsoi, T. Moriyama, T. Ono, and Y. Tserkovnyak, Rev. Mod. Phys. **90**, 015005 (2018).

[2] P. Wadley, B. Howells, J. Železný, C. Andrews, V. Hills, R.P. Campion, V. Novák, K. Olejník, F. Maccherozzi, S.S. Dhesi, S.Y. Martin, T. Wagner, J. Wunderlich, F. Freimuth, Y. Mokrousov, J. Kuneš, J.S. Chauhan, M.J. Grzybowski, A.W. Rushforth, K.W. Edmonds, B.L. Gallagher, and T. Jungwirth, Science **351**, 587 (2016).

[3] S.Y. Bodnar, L. Šmejkal, I. Turek, T. Jungwirth, O. Gomonay, J. Sinova, A.A. Sapozhnik, H.-J. Elmers, M. Kläui, and M. Jourdan, Nat Commun **9**, 1 (2018).

[4] X.Z. Chen, R. Zarzuela, J. Zhang, C. Song, X.F. Zhou, G.Y. Shi, F. Li, H.A. Zhou, W.J. Jiang, F. Pan, and Y. Tserkovnyak, Phys. Rev. Lett. **120**, 207204 (2018).

[5] T. Moriyama, K. Oda, T. Ohkochi, M. Kimata, and T. Ono, Sci Rep **8**, 1 (2018).

[6] I. Gray, T. Moriyama, N. Sivadas, G.M. Stiehl, J.T. Heron, R. Need, B.J. Kirby, D.H. Low, K.C. Nowack, D.G. Schlom, D.C. Ralph, T. Ono, and G.D. Fuchs, Phys. Rev. X **9**, 041016 (2019).

[7] L. Baldrati, O. Gomonay, A. Ross, M. Filianina, R. Lebrun, R. Ramos, C. Leveille, F. Fuhrmann, T.R. Forrest, F. Maccherozzi, S. Valencia, F. Kronast, E. Saitoh, J. Sinova, and M. Kläui, Phys. Rev. Lett. **123**, 177201 (2019).

[8] P. Zhang, J. Finley, T. Safi, and L. Liu, Phys. Rev. Lett. **123**, 247206 (2019).

[9] Y. Cheng, S. Yu, M. Zhu, J. Hwang, and F. Yang, Phys. Rev. Lett. **124**, 027202 (2020).

[10] L. Baldrati, C. Schmitt, O. Gomonay, R. Lebrun, R. Ramos, E. Saitoh, J. Sinova, and M. Kläui, ArXiv:2003.05923 [Cond-Mat] (2020).

[11] H. Nakayama, M. Althammer, Y.-T. Chen, K. Uchida, Y. Kajiwara, D. Kikuchi, T. Ohtani, S. Geprägs, M. Opel, S. Takahashi, R. Gross, G.E.W. Bauer, S.T.B. Goennenwein, and E. Saitoh, Phys. Rev. Lett. **110**, 206601 (2013).

[12] J. Fischer, O. Gomonay, R. Schlitz, K. Ganzhorn, N. Vlietstra, M. Althammer, H. Huebl, M. Opel, R. Gross, S.T.B. Goennenwein, and S. Geprägs, Phys. Rev. B **97**, 014417 (2018).

[13] L. Baldrati, A. Ross, T. Niizeki, C. Schneider, R. Ramos, J. Cramer, O. Gomonay, M. Filianina, T. Savchenko, D. Heinze, A. Kleibert, E. Saitoh, J. Sinova, and M. Kläui, Phys. Rev. B **98**, 024422 (2018).

[14] S. Geprägs, M. Opel, J. Fischer, O. Gomonay, P. Schwenke, M. Althammer, H. Huebl, and R. Gross, ArXiv:2004.02639 [Cond-Mat] (2020).

[15] A. Churikova, D. Bono, B. Neltner, A. Wittmann, L. Scipioni, A. Shepard, T. Newhouse-Illige, J. Greer, and G.S.D. Beach, Appl. Phys. Lett. **116**, 022410 (2020).

[16] T. Matalla-Wagner, J.-M. Schmalhorst, G. Reiss, N. Tamura, and M. Meinert, ArXiv:1910.08576 [Cond-Mat] (2019).

[17] C.C. Chiang, S.Y. Huang, D. Qu, P.H. Wu, and C.L. Chien, Phys. Rev. Lett. **123**, 227203 (2019).

[18] P. Wadley, S. Reimers, M.J. Grzybowski, C. Andrews, M. Wang, J.S. Chauhan, B.L. Gallagher, R.P. Campion, K.W. Edmonds, S.S. Dhesi, F. Maccherozzi, V. Novak, J. Wunderlich, and T. Jungwirth, Nature Nanotech **13**, 362 (2018).

[19] S.-W. Cheong, M. Fiebig, W. Wu, L. Chapon, and V. Kiryukhin, Npj Quantum Mater. **5**, 1 (2020).

[20] J. Xu, C. Zhou, M. Jia, D. Shi, C. Liu, H. Chen, G. Chen, G. Zhang, Y. Liang, J. Li, W. Zhang, and Y. Wu, Phys. Rev. B **100**, 134413 (2019).

[21] J. Xu, H. Chen, C. Zhou, D. Shi, G. Chen, and Y. Wu, ArXiv:2003.12768 [Cond-Mat] (2020).

[22] W.L. Roth, Journal of Applied Physics **31**, 2000 (1960).

[23] G.R. Hoogeboom, A. Aqeel, T. Kuschel, T.T.M. Palstra, and B.J. van Wees, Appl. Phys. Lett. **111**, 052409 (2017).

[24] D. Alders, L.H. Tjeng, F.C. Voogt, T. Hibma, G.A. Sawatzky, C.T. Chen, J. Vogel, M. Sacchi, and S. Iacobucci, Phys. Rev. B **57**, 11623 (1998).

[25] P. Němec, M. Fiebig, T. Kampfrath, and A.V. Kimel, Nature Phys **14**, 229 (2018).

# SUPPLEMENTARY MATERIAL

## S.1 Switched area detection

To give an example of the procedure used to extract the switched domain areas, we include in Fig. S1 the difference images corresponding to the switched area data plotted in Fig. 2(b), calculated as the contrast difference of domain images taken before and after the application of current pulses. By defining a threshold, the red bordered area and the corresponding number of switched pixels can be determined, which serve as a quantification of magnetic switching. A large change after the first pulse, quick saturation with only minor changes after the fifth pulse and a final stable domain state are clearly visible.

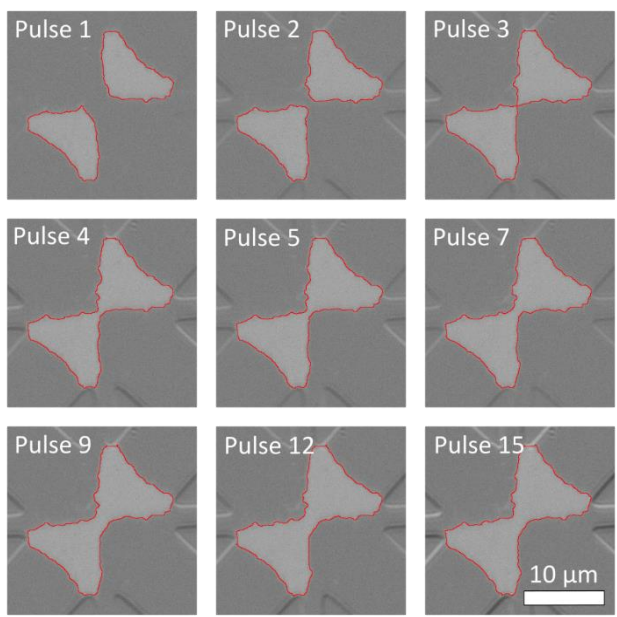

Fig. S1: Difference images corresponding to the switched area data in Fig. 2(b). The images show the difference of the domain structure after the application of $N$ +45° pulses ($j_P = 1.30 \times 10^{12}\ A\ m^{-2}$) with respect to the initial domain state (−45° pulse of $j_P = 1.50 \times 10^{12}\ A\ m^{-2}$, domain structure compares to Fig. 1(c)). The number of applied pulses $N$ is indicated in each panel. The scalebar provides the scale of all domain images.

## S.2 Transverse resistance signal of single switching measurement (as-measured)

The full set of electrical measurements including the $j_P = 1.30 \times 10^{12} \, A \, m^{-2}$ data plotted in Fig. 2(b) and corresponding to the modified data in Fig. 3(a) is shown in Fig. S2. We also show that the analysis has limitations: the data for high current densities $j_P > 1.425 \times 10^{12} \, A \, m^{-2}$ tends to deviate from the linear behavior and is therefore not included in the plots shown in Fig. 3.

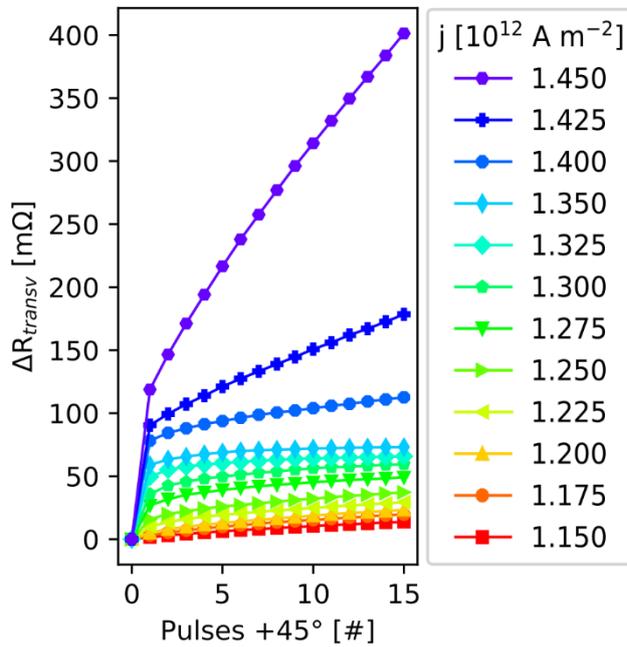

Fig. S2: Transverse resistance changes during the application of 15 pulses in $+45°$ direction after an initialization with a $-45°$ pulse of $j_P = 1.50 \times 10^{12} \, A \, m^{-2}$.